\documentclass[12pt]{iopart}
\usepackage{iopams}
\usepackage{graphicx}
\def\be{\begin{equation}}
\def\ee{\end{equation}}
\def\bea{\begin{eqnarray}}
\def\eea{\end{eqnarray}}
\newcommand{\ket}[1]{|#1\rangle}
\newcommand{\bra}[1]{\langle#1|}

\begin{document}

\title{Bipartite Entanglement in Continuous-Variable Cluster States}

\author{Hugo Cable}
\address{Centre for Quantum Technologies, National University of Singapore, 3 Science Drive 2, Singapore 117543}
\ead{cqthvc@nus.edu.sg}

\author{Daniel E. Browne}
\address{Department of Physics and Astronomy, University College London, Gower Street, London WC1E 6BT, United Kingdom}
\address{Centre for Quantum Technologies, National University of Singapore, 3 Science Drive 2, Singapore 117543}
\ead{d.browne@ucl.ac.uk}

\begin{abstract}
We present a study of the entanglement properties of Gaussian cluster states, proposed as a universal resource for continuous-variable quantum computing.  A central aim is to compare mathematically-idealized cluster states defined using quadrature eigenstates, which  have infinite squeezing and cannot exist in nature, with Gaussian approximations which are experimentally accessible.  Adopting widely-used definitions, we first review the key concepts, by analysing a process of teleportation along a continuous-variable quantum wire in the language of matrix product states.  Next we consider the bipartite entanglement properties of the wire, providing analytic results.  We proceed to grid cluster states, which are universal for the qubit case.  To extend our analysis of the bipartite entanglement, we adopt the entropic-entanglement width, a specialized entanglement measure introduced recently by Van den Nest M et al.\ 2006 {\it Phys.\ Rev.\ Lett.} {\bf 97} 150504, adapting their definition to the continuous-variable context.  Finally we add the effects of photonic loss, extending our arguments to mixed states. Cumulatively our results point to key differences in the properties of idealized and Gaussian cluster states.  Even modest loss rates are found to strongly limit the amount of entanglement.  We discuss the implications for the potential of continuous-variable analogues of measurement-based quantum computation.
\end{abstract}


\maketitle

\section{Introduction}

Continuous-variable (CV) quantum computing offers a promising route to optical quantum computing, being naturally suited to common experimental techniques using squeezed light and homodyne detection \cite{Braunstein05}.  CV Gaussian cluster states have been proposed as a resource for universal quantum computation using the one-way paradigm \cite{Raussendorf01,InitialProposalsCVCluster}.  Several theoretical proposals address the problems of efficient generation of these cluster states as well as performing the necessary local operations \cite{Gu09,GeneratingCVclusters}, and some proof-of-principle experiments have been reported \cite{CVClusterExperiments}.  These constitute part of a programme of adapting techniques from the context of discrete-variable (DV) -- primarily qubit-based -- quantum computing to phase space defined for infinite-dimensional modes, examples including quantum teleportation and other standard quantum-information protocols, and the matrix product state (MPS) representation \cite{GMPSPEPS}.

Often concepts are generalized from the qubit case to phase space by reference to some limit of infinite squeezing. Although such a mathematical procedure is intuitively appealing, the arguments typically used are not exact for physical states, since finitely-squeezed approximations of non-physical infinitely-squeezed states may demonstrate behaviour departing rapidly from the ideal.  For example, this can lead to cumulative errors in a computation \cite{Ohliger10}.  Strong squeezing is difficult to achieve in practice, although a high-squeezing factor was reported recently in \cite{Vahlbruch08}.  Squeezed states can be degraded by photonic loss and are commonly considered to be fragile, although this depends on the specific task and figure of merit \cite{SqueezingAndNoise}.  Together these issues place demanding requirements on error mitigation and fault tolerance strategies for implementing CV quantum computation.
An important line of investigation is reevaluate the usefulness of Gaussian cluster states as resource states when realistic assumptions on finite squeezing are applied.  To this end, the scaling of localisable entanglement was considered in \cite{Ohliger10}, the local complementation rule in \cite{Zhang08} and a CV implementation of the Deutsch-Jozsa algorithm in \cite{Adcock09}.

In this work, we consider several bipartite entanglement properties of Gaussian cluster states, making comparisons with their discrete analogues.  As an application, we consider the recently-introduced entropic-entanglement width (EW) measure, which provides a necessary condition for universality for families of resource states \cite{VandenNest07}.  We begin in section~\ref{sec:Basics} by introducing idealized CV cluster states, and their approximations using Gaussian states.  An intuitive demonstration of the problems of finite squeezing is presented by looking at a process of teleportation.  In section~\ref{sec:QWireBips}, we look at the entropic entanglement for all possible bipartitions of a quantum wire, focusing on the case of pure states.  The results of this are then applied, in section~\ref{sec:EWGrid}, to approximating the EW for a Gaussian grid cluster state, the qubit analogue of which is a universal resource for DV quantum computation.   In section~\ref{sec:Loss} we present numerical results on the effects of photon loss on the bipartite entanglement, adopting the logarithmic negativity as an entanglement monotone suitable for the mixed-state case. Finally, in section~\ref{sec:Conclusions} we discuss the relevance of these results for the prospects of scalable CV measurement-based quantum computation.

\section{Idealized CV and Gaussian cluster states and the effects of finite squeezing on coherent information transport}
\label{sec:Basics}

Let us first recall a widely-used definition of (idealized) CV cluster states \cite{InitialProposalsCVCluster,Gu09}.\footnote{An alternative approach might use a qubit encoding of computational states into the infinite-dimensional modes, as discussed in \cite{Gottesman01}.  An implementation using common-optical techniques was recently proposed \cite{Vasconcelos10}.  We do not pursue such encoding schemes here.}  For each mode labelled $i$ (termed here a ``qumode''), with annihilation operator $\hat{a}_i$, we work in a convention for which
 $\left[ \hat{a}_i,\hat{a}_i^{\dag }\right]\!=\!1$, $\hat{X}_i\!=\!\left(\hat{a}_i^{\dag }\!+\!\hat{a}_i\right)/\sqrt{2}$ and $\hat{P}_i\!=\!i\left( \hat{a}^{\dag }_i\!-\!\hat{a}_i\right)/\sqrt{2}$.  Given a graph with $N$ vertices, for which the neighbourhood of vertex $r$ is denoted $n(r)$, the corresponding CV cluster state is,
\begin{equation}
\label{eq:IdealCVCluster}
 \left\vert{\rm CV cluster}\right\rangle_{\infty}\!=\!\prod_{r=1}^{N}\prod_{s\in n(r)\vert s>r}\hat{U}_{CZ}^{\left( r,s\right) }\left\vert p=0\right\rangle _{1}\cdots \left\vert p=0\right\rangle _{N},
 \end{equation}
where the basis states are momentum-quadrature eigenstates satisfying $\hat{P}\ket{p}\!=\!p \ket{p}$ (for further explanation see \cite{BarnettRadmore}), and CV controlled-Z operations acting on pairs of modes are defined by $\hat{U}_{CZ}^{\left( r,s\right) }\!=\!\exp \left( i\hat{X}_{r}\hat{X}_{s}\right) $.

To explain the analogy with conventional quantum information processing using qubit cluster states, we consider as an example a process of ``one-bit teleportation'' \cite{Zhou00} along a computational quantum wire, and apply the MPS representation as in \cite{Gross08}.  The MPS formalism provides, for any choice of physical computational device,  a natural representation of the information-processing implemented by local measurements acting on the underlying computational resource states.  Accordingly, a qubit wire of $N$ sites, with initial input state $\ket{\psi_{\rm in}}$ (possibly unknown) defines a corresponding ``correlation space'' and matrices $A^{(\cdot,\cdot)}[\cdot]$ acting on it as:
\bea
\left\vert \psi _{\rm wire}\right\rangle
&=&\hat{U}_{\rm CZ}^{\left( N,N-1\right) }\cdots \hat{U}_{\rm CZ}^{\left( 2,1\right) }\left\vert \psi _{\rm in}\right\rangle _{1}\bigotimes_{j=2}^{N}|+\rangle_j \nonumber \\
&=&\sum_{\{x_{j}\}}\Big( \left\langle x_{N}\right\vert _{N}A^{\left( N,N-1\right) }\left[ x_{N-1}\right] \cdots A^{\left( 2,1\right) }\left[ x_{1}\right] \left\vert \psi _{\rm in}\right\rangle _{1}\Big) \bigotimes_{j=1}^{N} \left\vert x_{j}\right\rangle _{j},
\eea
where the $x_r$ label the computational basis states with values 0 and 1, $\left\vert \pm \right\rangle\!=\!\left( \left\vert 0\right\rangle \pm \left\vert 1\right\rangle \right) /\sqrt{2}$, and controlled-Z operations are defined by $\hat{U}_{\rm CZ}\!=\!\ket{00}\!\bra{00}\!+\!\ket{01}\!\bra{01}\!+\!\ket{10}\!\bra{10}\!-\!\ket{11}\!\bra{11}$.
Writing $\hat{U}_{\rm CZ}^{\left( r+1,r\right) }\left\vert +\right\rangle _{r+1}=\left\vert 0\right\rangle _{r}A^{\left( r+1,r\right) }\left[ 0\right] +\left\vert 1\right\rangle _{r}A^{\left( r+1,r\right) }\left[ 1\right]$, we see that MPS matrices acting between sites $r$ and $r\!+\!1$ are given by
$A^{\left( r+1,r\right) }\left[ 0\right]\!=\!\left\vert +\right\rangle _{r+1}\left\langle 0\right\vert _{r}$
and $A^{\left( r+1,r\right) }\left[ 1\right]\!=\!\left\vert -\right\rangle _{r+1}\left\langle 1\right\vert _{r}$.

Suppose we choose another basis labelled $\{\ket{\phi_r}\}$ and perform a measurement in this basis on site $r$, leading to projection  $\ket{\phi_r}\!\bra{\phi_r}_r$, the corresponding MPS matrix is given by $A^{\left( r+1,r\right) }\left[ \phi _{r}\right] \!=\!\sum_{x_r}\left\langle \phi _{r}\right\vert _{r}\left\vert x_{r}\right\rangle _{r}A^{\left( r+1,r\right) }\left[ x_{r}\right] $ (as follows directly from the definition of correlation space given above.)  To implement the paradigm of measurement-based quantum computing, we need to choose a measurement basis so that the
$A^{(r+1,r)}[\phi_{r}]$ are unitary (for example $A[0,1]$ are only rank 1).  Choosing the customary measurement basis $\{\ket{\pm}\}$ we find:
\bea
\label{eq:Site2SitePM}
A^{\left( r+1,r\right) }\left[ +_{r}\right] &=&\left( \left\vert +\right\rangle _{r+1}\left\langle 0\right\vert _{r}+\left\vert -\right\rangle _{r+1}\left\langle 1\right\vert _{r}\right) /\sqrt{2} \nonumber \\
A^{\left( r+1,r\right) }\left[ -_{r}\right] &=&\left( \left\vert +\right\rangle _{r+1}\left\langle 0\right\vert _{r}-\left\vert -\right\rangle _{r+1}\left\langle 1\right\vert _{r}\right) /\sqrt{2}.
\eea
These matrices are unitary (up to a normalization factor), involving a change-of-basis transformation (a Hadamard matrix) and, in the second case, an additional initial Pauli-$Z$ operation.  Hence measurements in the $\{\ket{\pm}\}$ basis drive unitary transport of $\ket{\psi_{\rm in}}$ along the wire, and the randomness of the measurement outcomes can be corrected by a local unitary operation at every step (thus implementing deterministic computation using feedforward control). Measurements along other axes in the x-y plane of the Bloch sphere allow one to generate any SU(2) logical gate, from which a fully universal set can be achieved by addition of an entangling gate.

We proceed now to make the analogy for transporting an initial unknown ``qumode'' state $\ket{\psi_{\rm in}}$ along an idealized CV quantum wire, using the language of generalized functions.  Defining the wire using (\ref{eq:IdealCVCluster}), we identify the computational basis states $\ket{x_r}_r$ as position-quadrature eigenstates (for which $\hat{X}\ket{X\!=\!x}\!=\!x \ket{X\!=\!x}$, the $x$-labels are now continuous, $\langle X\!=\!x \vert X\!=\!y\rangle\!=\!\delta \left( x\!-\!y\right) $ and $\int_{-\infty }^{\infty }dx\left\vert X\!=\!x\right\rangle \!\!\left\langle X\!=\!x\right\vert\!=\!I$).  The momentum quadrature (generalized) basis now plays the role of the $\ket{\pm}$ basis for the qubit case, and the quadrature bases are related by $\langle X\!=\!x \vert P\!=\!p \rangle\!=\!\exp\left(ixp\right)/\sqrt{2\pi }$.  Applying the MPS formalism for the CV case, the $A[\cdot]$ matrices in the computational basis are given by:
$A^{(r\!+\!1,r)}[x_r]=\left\vert P\!=\!x_{r}\right\rangle _{r\!+\!1}\left\langle X\!=\!x_r\right\vert_r$.  For measurements in the momentum basis (corresponding to homodyne detection):
\be
\label{eq:IdealCVMPS}
A^{(r+1,r)}_\infty\left[ -p_{r}\right] =\left( 2\pi \right) ^{-1/2}\left\{ \int dx_{r}\left\vert P\!=\!x_{r}\right\rangle _{r+1}\left\langle X\!=\!x_{r}\right\vert _{r}\right\} \exp \left( ip_{r}\hat{X}_{r}\right),
\ee
which is unitary up to a normalization factor.  Analogy is achieved in the ideal case with the matrices $A[\pm]$ for the qubit case: $\exp \left( i s\hat{X}\right) $ induces a translation $\ket{P}\mapsto\ket{P\!+\!s}$ and plays a similar role to Pauli-$Z$ (likewise $\exp (-it\hat{P})$ induces a translation $\ket{X}\mapsto\ket{X\!+\!t})$ and can be identified with Pauli-$X$); the Fourier transform operation corresponds to the ``always-on'' Hadamard operator.

Going further, the Weyl-Heisenberg group of displacement operators (acting on a qumode) might be identified with the Pauli group (acting on a qubit).  Gaussian operations, acting by conjugacy, transform each quadrature operator to a linear combination of the quadrature operators (possibly with an additional translation and all coefficients being real).  They therefore transform every displacement operator to another displacement operator, and stand in the same relation to the Weyl-Heisenberg group as the Clifford group does to the Pauli group \cite{Bartlett02}.  A non-Gaussian interaction characterized by a third-order Hamiltonian, or a process of photocounting \cite{Gu09}, can achieve universal CV quantum computation in the sense set out in \cite{Braunstein05}, in a similar way as a non-Clifford operation such as a single-qubit $\pi/8$-phase gate is required for universal quantum computing with qubit systems.

Having made these analogies, a more realistic assumption is to replace the quadrature eigenstates for the CV cluster state with Gaussian states having finite squeezing.  (We retain projections in the momentum-quadrature basis for the detection component, as these can be realised to good approximation by homodyne detection.)  For each mode in (\ref{eq:IdealCVCluster}) replace
$\ket{P\!=\!0}\mapsto\hat{S}^{\left( 1\right) }\left( \zeta \right)\ket{\rm vac}$, where the unitary one-mode squeezing operator is defined by
$\hat{S}^{\left( 1\right) }\left( \zeta \right) \!=\!\exp \left[ \left( -\zeta \hat{a}^{\dag 2}\!+\!\zeta ^{\ast }\hat{a}^{2}\right) /2\right]$
and $\zeta\!=\!\left\vert \zeta \right\vert \exp \left( i\varphi \right) $.  The following relations encapsulate the effect of $\hat{S}^{(1)}(\zeta)$ in phase space.
\bea
\hat{S}^{\left( 1\right) \dag } \hat{X}\hat{S}^{\left( 1\right) }
\!=\!\left[ \cosh \left( \left\vert \zeta \right\vert \right) \!-\!\cos \left( \varphi \right) \sinh \left( \left\vert \zeta \right\vert \right) \right] \hat{X}-\sin \left( \varphi \right) \sinh \left( \left\vert \zeta \right\vert \right) \hat{P} \nonumber \\
\hat{S}^{\left( 1\right) \dag } \hat{P}\hat{S}^{\left( 1\right) }
\!=\!-\sin \left( \varphi \right) \sinh \left( \left\vert \zeta \right\vert \right) \hat{X}+\left[ \cosh \left( \left\vert \zeta \right\vert \right) \!+\!\cos \left( \varphi \right) \sinh \left( \left\vert \zeta \right\vert \right) \right] \hat{P}.
\eea
Setting $\zeta=-\vert \zeta \vert$ achieves a squeezing for momentum, and an anti-squeezing for position.  The squeezed vacuum has only even photon number components, and hence the expectation value for the quadrature operator in any phase-space direction is zero.

It is now possible to write down directly MPS matrices $A[\cdot]$ corresponding to (\ref{eq:IdealCVMPS}) above, for physical Gaussian cluster states.  Expanding the squeezed vacuum in terms of momentum-quadrature eigenstates,
$
\int_{-\infty }^{\infty }dp|P\!=\!p\rangle \! \left\langle P\!=\!p\right\vert \left\vert -\left\vert \zeta \right\vert \right\rangle =\left[ 2\pi \left( \Delta ^{2}p\right) \right] ^{-1/4}\int_{-\infty }^{\infty }dp\exp \left[ -\frac{p^{2}}{4\left( \Delta ^{2}p\right) }\right] |P\!=\!p\rangle,
$
having a Gaussian distribution about $P\!=\!0$ with $\Delta ^{2}p=\left[ \exp \left( -2\left\vert \zeta \right\vert \right) \right] /2$. Then,
\be
A_{\rm physical}^{\left( r+1,r\right) }[-p_{r}]
\propto \left\{
\int_{-\infty }^{\infty }dp\exp \left[ -\frac{p^{2}}{4\left( \Delta ^{2}p\right) }\right]
\exp \left( ip\hat{X}_{r+1}\right) \right\}
 A_{\infty }^{\left( r+1,r\right) }[-p_{r}].
\ee
We see that the effect of finite squeezing is to introduce an additional smearing in the computational basis (which has been mapped to the momentum basis after the action of the measurement on qumode $r$).  The $A[\cdot]$ matrix no longer acts as a unitary map along the wire, and the errors arising from measurement are not simply correctable by feedforward.  Every attempt to transport the original qumode another step along the Gaussian wire will introduce a further spreading.  Hence the effect of finite squeezing is very detrimental to accurate coherent information transport, a point previously made in \cite{Gu09,Ohliger10}.  We now proceed to investigate the computation power of Gaussian cluster states from another angle - namely their bipartite entanglement properties.

\section{Pure-state entanglement for bipartitions of the quantum wire}
\label{sec:QWireBips}

In this section, we begin our study of the bipartite entanglement properties of Gaussian cluster states, disregarding decoherence processes for the time being.  Since these quantum states are pure, we can adopt the entropic entanglement (EE) as a convenient measure; it is defined by $EE(\beta_1,\beta_2)\!=\!-\tr\left[ \rho _{\rm red}\log \left( \rho _{\rm red}\right) \right]$ \footnote{Following common conventions for EE, we use the natural logarithm for CV calculations, and base 2 for qubit calculations.}, where $\beta_1$ and $\beta_2$ denote a bipartitioning of the vertices, and $\rho _{\rm red}$ denotes either of the corresponding reduced density matrices.  Entanglement monotones are invariant under local unitary transformations, and so controlled-Z operations within $\beta_1$ or $\beta_2$  can safely be disregarded, along with qumode/qubit vertices which become disconnected this way.  Our basis for comparison is the EE properties for qubit cluster states, which can be determined for any given bipartition by simple counting arguments \cite{VandenNest06}.  More specifically, let $A$ denote the adjacency matrix of some finite graph (such that $A_{ij}$ is 1 if vertices $i$ and $j$ are joined by an edge and 0 otherwise), and $A^\prime$ denote the submatrix obtained by deleting from $A$ rows corresponding to the elements of $\beta_1$ and columns corresponding to $\beta_2$.  $A^\prime$ then encodes information about edges between $\beta_1$ and $\beta_2$, and $EE(\beta_1,\beta_2)$ is given by the binary matrix rank of $A^\prime$ (i.e. arithmetic modulo 2, and where base-2 is used for the definition of EE) \cite{Hein06}.  In a qubit cluster state, the contribution to $EE(\beta_1,\beta_2)$ from a particular vertex cannot be more than 1 whether it is singly or multiply bonded, since the maximum entropy of a qubit is 1.  In the CV case, however, there is no upper bound to the entropy of an individual qumode and the entropy of entanglement can exhibit a richer range of behaviour, as we shall see below.

For our CV calculations, we use phase-space methods \cite{PhaseSpace}, defined for $N$ modes as follows. The $2N$
canonical-coordinate operators and variables are denoted by column vectors
$\mathbf{\hat{O}}\!=\!\left( \hat{X}_{1};\hat{X}_{2};\cdots \hat{P}_{1};\hat{P}_{2}\cdots \right) $
and $\boldsymbol{\xi }=\left( X_1;X_2;\cdots P_{1};P_{2}\cdots \right) $.
Matrix $\Sigma $ is defined by $\left[ \hat{O}_{j},\hat{O}_{k}\right] =i\Sigma _{jk}$,
where $\left[ \hat{X},\hat{P}\right] =i$.
A (pure or mixed) state $\rho $ is called Gaussian if its characteristic
function, $\chi _{\rho }\left( \mathbf{\xi }\right) =\tr\left[ \rho
\exp \left( -i\mathbf{\xi }^{t}\Sigma \mathbf{\hat{O}}\right) \right] $, is
completely determined by its first and second moments and can be written in terms of a displacement vector $\mathbf{d}$ and covariance matrix $\Gamma $,
$\chi _{\rho }\left( \mathbf{\xi }\right) \!=\!\exp \left( -\frac{1}{4}\mathbf{\xi }^{t}\Sigma \Gamma \Sigma ^{t}
\mathbf{\xi }+i\mathbf{d}^{t}\Sigma \mathbf{\xi }\right) $,
where
$\tr\left( \rho\hat{O}_a\right)=d_a$,
and
$\Gamma_{ab}\!=\!
2\tr\left[ \rho
\left( \hat{O}_a-d_a \right) \left( \hat{O}_b-d_b \right) \right]
 -i\Sigma _{ab}$.
A further condition applies: $\Gamma $ represents a physical state if and only if $\Gamma \pm i\Sigma\geq 0$.
Given a unitary transformation $\hat{U}$ on the state space for which
$\hat{U}^\dag\hat{O}_{i}\hat{U}=\sum_{j}S_{ij}\hat{O}_{j}$, the matrix $S$ is a real and symplectic
(namely it satisfies the condition $S^{t}\Sigma S=\Sigma $).
The transformation of a Gaussian state $\rho \mapsto \hat{U}\rho \hat{U}^{\dag}$
is equivalent to the mappings $\Gamma \mapsto S\Gamma S^{t}$ and
$\mathbf{d}\mapsto S\mathbf{d}$.

For a Gaussian $N$-mode state, the EE can be evaluated directly from the covariance matrix.
Given a bipartition of the modes, the reduced state for one component is Gaussian, and
the corresponding reduced covariance matrix $\Gamma_{\rm red}$ is given by removing rows and columns
corresponding to the modes which are traced out.
Applying Williamson's theorem, which provides a normal-mode decomposition for
real, symmetric and positive-definite matrices \cite{Williamson36},
any covariance matrix $\Gamma$ can be diagonalised
via a  symplectic transformation.  The corresponding ``symplectic eigenvalues'' come in same-value pairs with positive values
$\lambda_1,\cdots,\lambda_N$ (where $\lambda_i\geq1$ for physical states).  The matrix $\Sigma \Gamma$ has eigenvalues
$\pm i\lambda _{1}$,$\cdots $,$\pm i\lambda _{N}$, and can be used to compute the symplectic spectrum.
The diagonalising symplectic transformation in phase-space corresponds to a unitary transformation on the state-space, relating
the Gaussian state to thermal states in $N$ independent modes.  These results allow a simple
formula for EE to be derived, from the symplectic spectrum $\lambda_{k}^{\rm red}$ of $\Gamma_{\rm red}$:
\be
\label{eq:GaussianEE}
EE=\sum_{k=1}^{N}\left[ \left( \frac{\lambda^{\rm red} _{k}+1}{2}\right)
\ln \left( \frac{\lambda^{\rm red}_k+1}{2}\right)
 -\left( \frac{\lambda^{\rm red}_k-1}{2}\right) \ln \left( \frac{\lambda^{\rm red} _k-1}{2}\right) \right].
\ee

We now focus on the bipartite entanglement behaviour of a quantum wire.  For any given bipartition, we disregard controlled-Z operations which do not contribute to the EE, and thereby break the wire into disconnected lengths of ``zigzig-type'' strings, for which alternate vertices switch bipartition component.  The simplest zigzag has only two vertices and a single bond.  Disconnected zigzags contribute additively to the total EE.  Hence the EE for all possible bipartitions can be understood by considering only zigzag bipartitions.  For a continuous zigzag bipartition of a $N$ qubit wire, the EE is easily seen to have value $(N-1)/2$ when $N$ is odd, and $N/2$ when $N$ is even.  We progress now to the case of Gaussian quantum wires defined as section~\ref{sec:Basics}.  We note that a related investigation is reported in \cite{Audenaert02}, which looks at the ground and thermal states of a closed harmonic chain with nearest-neighbour couplings.

We first transform the $N$-qumode Gaussian quantum wire by local unitary squeezing processes, absorbing the initial momentum-squeezing of each qumode into the CV controlled-Z operations:
\be
\label{eq:GWire}
\hat{S}_{\rm all}^{\dag }\left\vert {\rm G-wire}\right\rangle
\!=\!
\prod_{a\in 1}^{N-1}\left( \hat{S}_{\rm all}^{\dag }\hat{U}_{CZ}^{\left( a,a+1\right) }\hat{S}_{\rm all}\right) \left\vert 0\right\rangle,
\ee
where $\hat{S}_{\rm all}\!=\!\otimes _{r=1}^{N}\hat{S}\left( -\left\vert \zeta \right\vert \right)_r$
and
$\hat{S}_{\rm all}^{\dag }\hat{U}_{CZ}^{\left( a,b\right) }\hat{S}_{\rm all}\!=\!\exp \left( ie^{-2 \zeta }\hat{X}_a\hat{X}_b\right)$.
In other words, we can replace the single-mode squeezing and controlled-Z operators by a single two-mode squeezing operator, whose strength is given by the squeezing parameter $B\!=\!e^{-2 \zeta }$, solely dependent upon the degree of squeezing of the initial single-mode states. By the Baker-Hausdorff lemma, the symplectic matrix for two arbitrary modes labelled $i$ and $j$ corresponding to gate operation $\hat{U}^{(i,j)}(B)=\exp \left( iB\hat{X}_i\hat{X}_j\right)$ is,
\be
\label{eq:BondMat}
S(B) =
\left(
\begin{array}{cccc}
1 &	0 &	0 &	0 \\
0 &	1 &	0 &	0 \\
0 &	B &	1 &	0 \\
B &	0 &	0 &	1 \\
\end{array}
\right).
\ee

Let us begin by studying the simplest zigzag, namely two modes connected (across the bipartition) by a single bond. For state $\exp \left( iB\hat{X}_{1}\hat{X}_{2}\right) |0\rangle _1|0\rangle _2$, the symplectic
eigenvalue for the reduced density matrices is $\lambda\!=\!\sqrt{1+B^{2}}$, and there is one contribution to the EE in (\ref{eq:GaussianEE}).
Provided $\lambda$ is not close to $1$, a convenient approximate form for the entanglement entropy of this two-mode state (which upper bounds the exact expression, and converges rapidly to it with increasing $\lambda$) can be readily derived,
\bea
\label{eq:approxEE}
EE_{\rm S}&=&\left( \frac{\lambda\!+\!1}{2}\right) \ln \left( \frac{\lambda\!+\!1}{2}\right)
-\left( \frac{\lambda\!-\!1}{2}\right) \ln \left( \frac{\lambda\!-\!1}{2}\right) \nonumber \\
&\simeq& \ln \left( \frac{\lambda}{2}\right) +1. 	
\eea
Hence, in terms of the squeezing parameter $B$, the entanglement entropy for $B\gg1$ has the simple form $EE_{\rm S} \simeq \ln(B/2)+1$.

\begin{figure}[t]
\begin{center}
\includegraphics[width=3.7in]{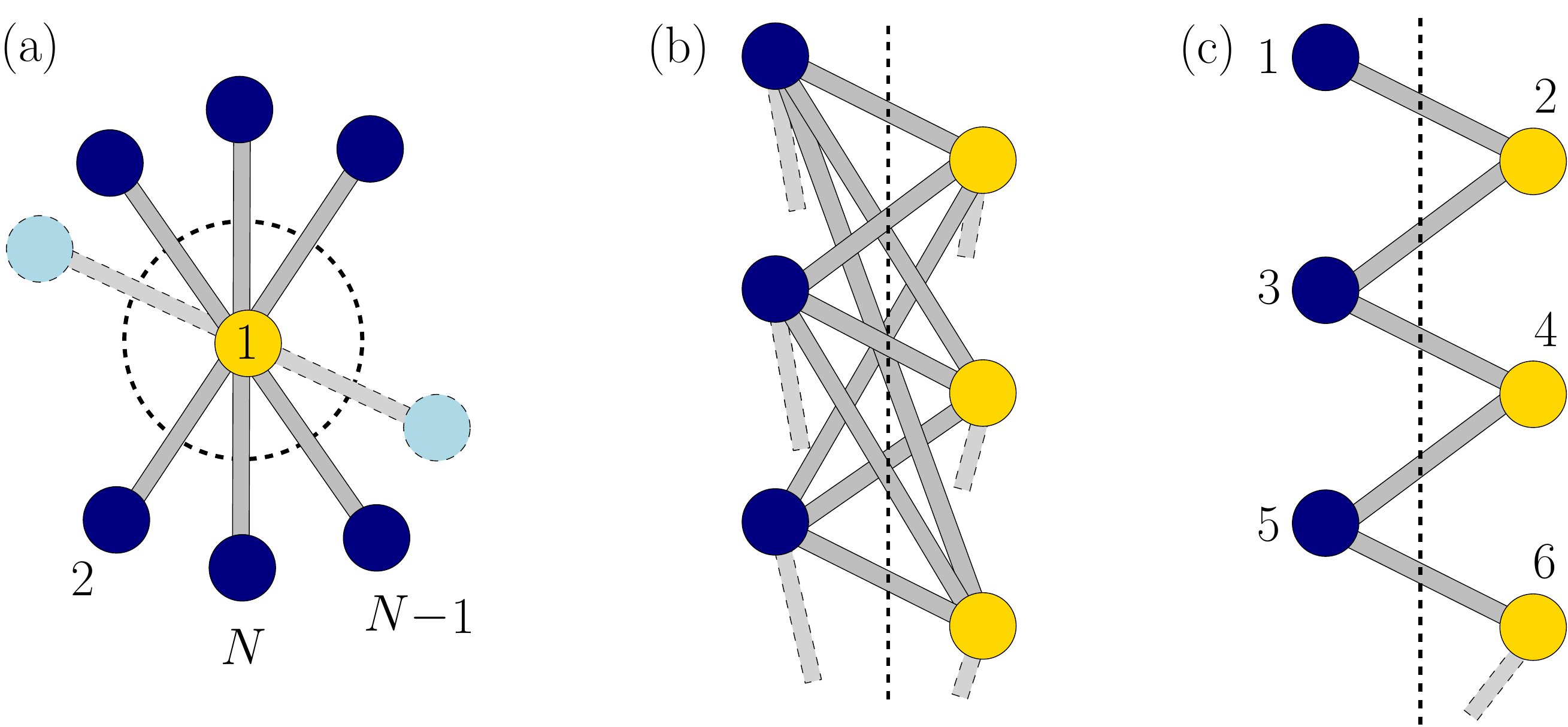}
\end{center}
\caption{\label{fig:SimpleGraphStates}Three classes of $N$-qumode Gaussian cluster states, generated by CV controlled-Z operations with parameter $B$.
(a) Star configuration with bipartition defined by central qumode 1; (b) completely-connected configuration bipartitioned with components of size $r$ and $N\!-\!r$ (controlled-Z operators that do not contribute to the EE have been omitted); (c) zigzag configuration bipartitioned into qumodes with odd or even labels.  For calculating the EE, cases (a) and (b) are equivalent to the bipartite entanglement for a pair of qumodes acted upon by a controlled-Z operation with $B\mapsto\sqrt{N\!-\!1}B$ and $B\mapsto{\sqrt{r}\sqrt{N\!-\!r}B}$ respectively.}
\end{figure}

We can calculate the entanglement entropy for certain related graph states, across appropriate bipartitions, by reducing them to the single bond case, but with a modified squeezing parameter.
Consider a star Gaussian cluster state with $N\!-\!1$ edges:
$\exp \left( iB\hat{X}_{1}\hat{X}_{2}\right) \exp \left( iB\hat{X}_{1}\hat{X}_{3}\right) \cdots \exp \left( iB\hat{X}_{1}\hat{X}_{N}\right) |0\rangle _{1}|0\rangle _{2}\cdots |0\rangle _{N}$ ($1$ labels the centre mode).
Taking the bipartition of mode $1$ versus all other modes, as illustrated in figure \ref{fig:SimpleGraphStates}(a), we can transform this state via a local unitary (with respect to the bipartition) to a two-mode single bond state. We achieve this via the mapping $(\hat{X}_{2}+\hat{X}_{3}\cdots \hat{X}_{N})\mapsto \sqrt{N-1}\hat{X}_{2}^{\prime }$. This can be achieved using a $N\!-\!1$ mode unitary beam-splitter network implementing the finite Fourier transform (defined by the linear transformation $\hat{a}_{s}\mapsto \sum_{t}F_{st}\hat{a}_{t}$ with $\left( F\right) _{st}\!=\!\left[ \exp (i2\pi(s\!-\!1)(t\!-\!1)/(N\!-\!1))\right]/\sqrt{N\!-\!1}$) \cite{Reck94}.  Hence we see that the star state has the same EE as the single bond state with squeezing enhanced by a factor of $\sqrt{N\!-\!1}$. For initial squeezing $B$, the EE for the star state is approximately $EE_{\rm star} \simeq \ln\left(\sqrt{N\!-\!1}B/2\right)+1$.

Another example is the completely-connected graph state, defined by
$\prod_{i\!=\!1}^{N}\prod_{j\!=\!i}^{N}\exp \left( iB\hat{X}_{i}\hat{X}_{j}\right) |0\rangle _{1}|0\rangle _{2}\cdots |0\rangle _{N}$, where a bonding operation has been applied once to every pair of modes. Given a bipartition of the modes into components with size $r$ and $N\!-\!r$, as illustrated in figure \ref{fig:SimpleGraphStates}(b), we can use a similar trick as above to map the state into a single bonded pair with increased bond strength $B\mapsto \sqrt{r}\sqrt{N-r}B$. The entanglement entropy in the strongly-squeezed limit is $EE_{\rm complete} \simeq \ln(\sqrt{r}\sqrt{N\!-\!r}B/2)+1$.  A maximizing bipartition then has components of equal size.  We have found that the scaling for the EE is logarithmic in the system size, both for the star and completely-connected configurations of the Gaussian cluster state.  Following similar arguments to \cite{VandenNest06}, we can immediately conclude that these two families of states do not constitute universal resources, for any value of the squeezing parameter $B$.

The qubit analogies for these states behave rather differently.  In particular, the $N$-qubit cluster states in the star and completely-connected configurations share the same entanglement properties.  This is true since a star graph can be converted to the completely-connected graph by the local complementation rule (applied to the centre qubit), which implies the existence of a local Clifford operation transforming one into the other \cite{Hein06}.  Therefore, for every nontrivial bipartition of these qubit states, the EE has value 1.

We will now consider a wire of length $N$, initially in the vacuum state, subject to operations $\exp \left( iB\hat{X}_{j}\hat{X}_{j+1}\right)$ on pairs of qumodes $(1,2)$,$(2,3)$,$\cdots$,$(N-1,N)$. We bipartition the wire as a zigzag with all odd-numbered modes constituting one half of the chosen bipartition (as illustrated in figure \ref{fig:SimpleGraphStates}(c)).  The reduced covariance matrix for the even-numbered modes is of the block form,
\be
\Gamma_{\rm zigzag}^{\rm red}\left( N\right) =
\left(
\begin{array}{cc}
I_{N_0} & 0 \\
0 &	T_{N_0} \\
\end{array}
\right),
\ee
where the blocks have dimension $N_0\!=\!\frac{N\!-\!1}{2}$ by $\frac{N\!-\!1}{2}$ when $N$ is odd, and $N_0\!=\!\frac{N}{2}$ by $\frac{N}{2}$ when $N$ is even. $I_{N_0}$ denotes the identity matrix, and $T_{N_0}$ is a tridiagonal matrix.  When $N$ is odd, the leading diagonal of $T_{N_0}$ has constant value $(1\!+\!2B^2)$, and the adjacent diagonals have value $B^2$ everywhere.  The $N$ even case is similar, with the bottom right corner element altered to $\left( 1\!+\!B^{2}\right) $.   For the symplectic spectrum corresponding to the zigzag bipartition, we need the eigenvalues $\Lambda_k$ of $\Sigma \Gamma_{\rm zigzag}^{\rm red}\left( N\right)$.  These also solve the eigenvalue equation
${T}_{N_0}{v}\!=\!-\Lambda_k^2{v}$.  The spectrum of ${T}_{N_{0}}$ is easily computed for odd and even $N$ cases \cite{PatternedMatrices}, and we find,
\be
\label{eq:ZigzagSympSpectrum}
\Lambda_k=\pm i\sqrt{1\!+\!2B^{2}\left(1+ \cos \left( \frac{2k\pi }{N\!+\!1}\right)\right) },
\ee
where $k\!=\!1\cdots N_{0}$; the EE can be computed from the absolute values using (\ref{eq:GaussianEE}). This expression appears rather complicated, but we can derive a much simpler form valid in the limit of large squeezing and many modes. We shall focus only on the case where $N$ is odd since it will play an important role in calculating the entropic-entanglement width for a square-lattice Gaussian cluster state later in this article.

We begin by taking the high-squeezing approximation (\ref{eq:approxEE}) for every term in (\ref{eq:GaussianEE}).
\begin{eqnarray}
EE_{\textrm{zigzag}}(N,B)&\simeq&\sum_{k=1}^{N_0}\left[ \ln(\lambda_k)\!+\!1\!-\!\ln(2)\right] \nonumber\\
&=&\frac{1}{2}\sum_{k=1}^{N_0} \ln\left[1\!+\!2B^2\left(1\!+\!\cos \left( \frac{2k\pi }{N\!+\!1}\right)\right)\right]\nonumber \\
&+& N_0[1\!-\!\ln(2)]. \nonumber
\end{eqnarray}
In the limit that $N_0$ is large, we can approximate this sum by an integral to obtain,
\[
EE_{\textrm{zigzag}}(N,B)\simeq \frac{N_0}{2\pi} \int_0^\pi \ln[1\,+\,2B^2(1\,+\,\cos(x))]dx\,+\,N_0\left[1\!-\!\ln(2)\right],
\]
and upon integrating \cite{HandBookMathematics},
\be
\label{eq:ZZApproxWithConstant}
EE_{\textrm{zigzag}}(N,B)\simeq N_0\left[1\!-\!\frac{3}{2}\ln(2)\!+\!\ln\left(\sqrt{1\!+\!2B^2\!+\!\sqrt{1\!+\!4B^2}}\right)\right],
\ee
and hence,
\be
\label{eq:ZZApproxWithoutConstant}
EE_{\textrm{zigzag}}(N,B)\approx N_0\left[\ln\left(\sqrt{1\!+\!2B^2\!+\!\sqrt{1\!+\!4B^2}}\right)\right].
\ee
We find that (\ref{eq:ZZApproxWithoutConstant}) converges quickly to the value of the exact expression as $B$ and $N$ grow large.

\section{Entropic-entanglement width for the grid state}
\label{sec:EWGrid}

In this section we consider the entanglement properties of the family of Gaussian cluster states in the $l$-by-$l$--grid configuration, which is well known to be universal for DV quantum computing.  To do this we adopt the entropic-entanglement width (EW), introduced by Van den Nest et al.\ \cite{VandenNest07}.  The previous authors focused on the implications of the scaling of the EW, and provided a no-go result for (efficient) universality when the EW is bounded (or scales at most logarithmically) for a family of resource states.   In this work we extend the definition of EW, replacing qubits with qumodes, and we look primarily at qualitative differences in the behaviour of Gaussian cluster states compared to qubit cluster states.

\begin{figure}[t]
\begin{center}
\includegraphics[width=4.1in]{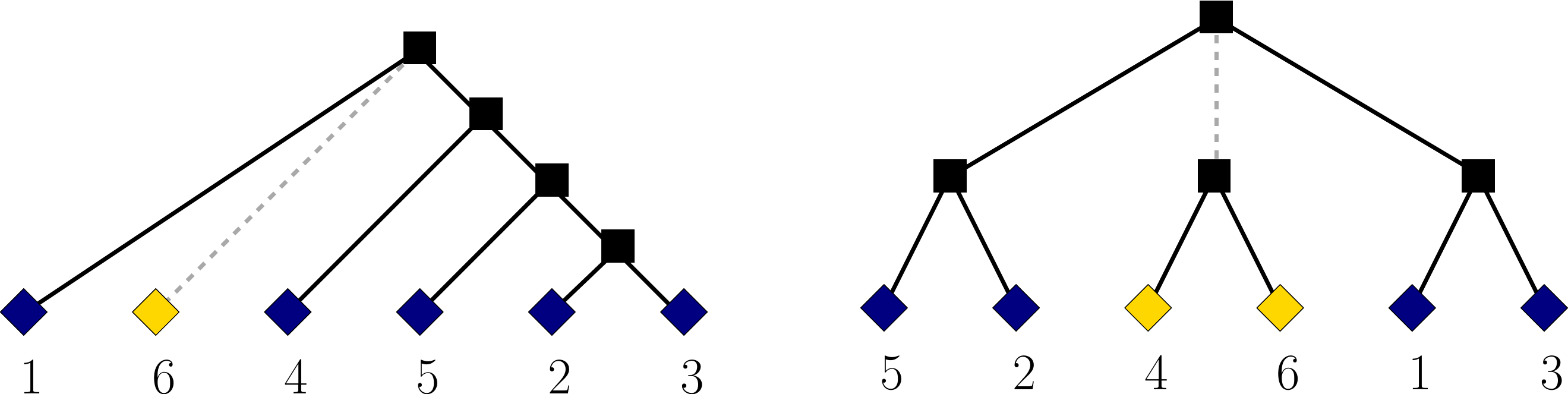}
\end{center}
\caption{\label{fig:ExplainingTrees} Two examples of decompositions, each defined by a subcubic tree (for which every vertex has one or three incident edges) and a bijection from the tree leaves to the qumodes.  Removing an edge defines a bipartition.  Vertices with only two incident edges are associated with only one bipartition, and can be removed by merging the edges.  Although not necessary for the definition of the EW, subcubic trees can be illustrated as rooted trees, for which an arbitrary non-leaf vertex has been designated the root, defining a partial ordering.}
\end{figure}
To begin, we introduce the idea of using a subcubic tree (a tree graph having one or three edges incident at each vertex) to induce a set of bipartitions of a collection of qumode (or qubit) labels.\footnote{The use of trees here is entirely unrelated to any graph appearing in the definition of the state.}  The situation is illustrated in figure \ref{fig:ExplainingTrees}.  A tree is an undirected graph in which any two vertices are connected by exactly one simple path.  It follows from this definition that any tree is connected, without cycles, and that the number of edges is one less than the number of vertices.  Vertices with only one incident edge are termed leaves.  Each leaf is identified with one qumode label, and a tree together with this labelling is termed a decomposition.  Given a decomposition $T$ for a $N$-qumode state $\ket{\psi}$, removal of an edge $e$ disconnects the tree, and bipartitions the set of qumodes $Q$ into subsets $\beta_e$ and $Q\setminus \beta_e$.  Each edge of $T$ therefore defines a value for the EE across the corresponding bipartition.  The maximum of these values for $T$ is termed the width of $T$.  The EW of $\ket{\psi}$ is then defined as the minimum width for all possible decompositions:
\be
\label{eq:DefEW}
EW(\ket{\psi})={\min_T} \left( {\max_{e\in T}EE}(\beta_e,Q\setminus \beta_e)\right).
\ee

We thus see that any valid decomposition can provide an upper bound to the EW, but finding the optimal decomposition (and hence a tight value for the EW) can be difficult. For qubit cluster states, the EE has the special form explained in section~\ref{sec:QWireBips}, and hence the EW \cite{VandenNest06} coincides with a graph width parameter called the rankwidth \cite{Oum06}.  For a one-dimensional cluster state (a ``wire'') the rankwidth is easily seen to have value 1.  For the $l$-by-$l$ grid, by consideration of the bipartition defined by the main diagonal, one can show that an upper bound to the rankwidth is $l\!-\!1$. Recently, by a series of clever arguments which exploit the symmetries of the square grid structure. it has been shown that this bound is tight and that the rankwidth of the grid is $l\!-\!1$ \cite{Jelinek10}.  To evaluate the difficulty of computing  exhaustively the EW for an arbitrary $N$-qumode state, we count the total number of possible decompositions (for $N=2,3$ there is only one possibility).  Suppose that $T_{N\!-\!1}$ is a decomposition for $(N\!-\!1)$ qumodes; it can be turned into a decomposition for $N$ qumodes by the following step: split an edge and insert a non-leaf vertex together with an adjacent leaf labelled $N$.  This step adds two vertices and two edges.  Furthermore, every edge of every $(N\!-\!1)$-qumode decomposition leads to a different $N$-qumode decomposition, and all possible $N$-qumode decompositions are generated by this method.  It follows that
there are $\left( 2N\!-\!5\right) \left( 2N\!-\!7\right) \cdots 1=\left( 2N\!-\!5\right) !/\left[2^{N\!-\!3}\left( N\!-\!3\right)!\right]$ total decompositions for a $N$-qumode state ($N\geq3$), and direct evaluation of the EW is unfeasible unless $N$ is small.

\begin{figure}[t]
\begin{center}
\includegraphics[width=4.3in]{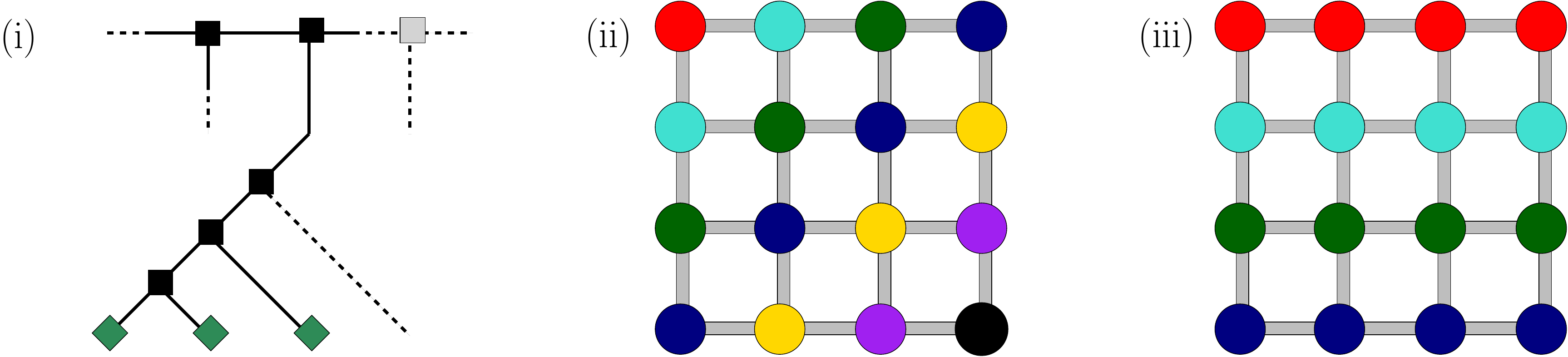}
\end{center}
\caption{\label{fig:WiringTogetherSubTrees} (i) A construction for building a decomposition from a chosen partition of graph vertices. A simple tree is constructed for each partition subset, and leaves are identified with the elements.  These trees are then ``strung together'' (suitably ordered) to form a subcubic tree and a decomposition for the entire graph.  For qubit cluster states defined on the $l$-by-$l$ grid, the width for the decomposition defined by diagonal subsets (ii) is $l\!-\!1$, and the decomposition is optimal.  For the rectangular case (iii), the width for the decomposition is $l$.  For Gaussian cluster states, the widths for (ii) and (iii) depend on the squeezing parameter.
}
\end{figure}
To address the problem of computing the EW for large grids, we select two candidate decompositions whose properties, in the case of qubit states, make them good candidates for optimal decompositions. Calculating the width of these decompositions provides upper bounds for the EW for the states. For small grids, we can then check the optimality of the decomposition via numerics.

The construction we use is illustrated in figure~\ref{fig:WiringTogetherSubTrees}(i).\footnote{Vit Jel{\'i}nek, private communication.}  Given a graph, we first partition the set of vertices $Q$ into subsets $\beta_1,\cdots,\beta_p$ (possibly of different sizes).  For each $\beta_i\!=\!\{v_{i1},v_{i2},\cdots\}$, we associate a simple tree.
Removing edges for this tree generates bipartitions $(\{v_{i1}\},Q\setminus\{v_{i1}\})$, $(\{v_{i1},v_{i2}\},Q\setminus\{v_{i1},v_{i2}\})$, and so on. We can then ``string'' these simple trees together in order, to yield a subcubic tree for the whole of $Q$.  Removing edges from the string generates bipartitions $(B_1,Q\setminus B_1)$, $\left(B_1\cup B_2,Q\setminus (B_1\cup B_2)\right)$, and so on.  The two decompositions we choose use diagonal bipartitions (figure~\ref{fig:WiringTogetherSubTrees}(ii)) and rectangular bipartitions (figure~\ref{fig:WiringTogetherSubTrees}(iii)), where the bipartitions are strung together from top left to bottom right, and from top to bottom respectively.

For the diagonal-type decomposition, the bipartition defined by the central diagonal is maximizing and determines the width of the decomposition (although it is not in general  the only maximizing partition). In both DV and CV cases, by applying local unitaries within the bipartitions, the calculation of the EE reduces to that for the zigzag-type bipartition, whose properties for Gaussian cluster states we analysed in section~\ref{sec:QWireBips}. For the case of a qubit cluster state, the width of the diagonal-type decomposition takes the optimal value, $l\!-\!1$.  For the rectangle-type decomposition, the maximizing bipartition will be given by a row of vertices away from the boundary of the grid. Applying  unitaries local with respect to the bi-partition we can remove all entangling ``bonds'' leaving $l$ separate and unentangled 3-mode wire states. For the case of a qubit cluster state, the width of the rectangle-type decomposition is immediately seen to be $l$, and the decomposition is suboptimal.

To calculate the width of these decompositions for a Gaussian cluster state of bond-strength $B$ defined on the $l$-by-$l$ grid, we can directly use results derived in section~\ref{sec:QWireBips}. First we shall consider the diagonal-type decomposition. Via application of local unitaries with respect to the bipartition defined by the central diagonal, the state can be reduced to a $(l-1)$ length zigzag cluster state with bond strength $B'=\sqrt{2}B$. Then, assuming $B$ is not small, we can directly use the approximate expressions (\ref{eq:ZZApproxWithConstant})  or more compactly (\ref{eq:ZZApproxWithoutConstant}) to achieve,
\be
\label{eq:AnalyDiag}
EE_{\rm diag}(l,B)\approx (l\!-\!1)\ln\left(\sqrt{1\!+\!4B^2\!+\!\sqrt{1\!+\!8B^2}}\right).
\ee

For the rectangle-type decomposition,  by the application of further local unitaries (as for the star graph in the section~\ref{sec:QWireBips}) we can reduce $l$ length-3 zigzags to $l$ singly-bonded pairs across the bipartition with bond strength $\sqrt{2}B$.  The corresponding EE is therefore $l$ times that of a simple pair with this bond strength.  We use (\ref{eq:approxEE}) with the single symplectic eigenvalue $\lambda=\sqrt{1\!+\!2B^2}$  to directly obtain the following expression, a good approximation when $\lambda$ is not close to $1$,
\bea
\label{eq:AnalyRect}
EE_{\rm rect}(l,B)&\simeq& l\left[ \ln(\sqrt{1\!+\!2B^2})\!+\!1\!-\!\ln(2)\right] \nonumber \\
&\approx& l\left[ \ln\left(\sqrt{1\!+\!2B^2}\right)\right].
\eea

\begin{figure}[t]
\begin{center}
\includegraphics[width=2in]{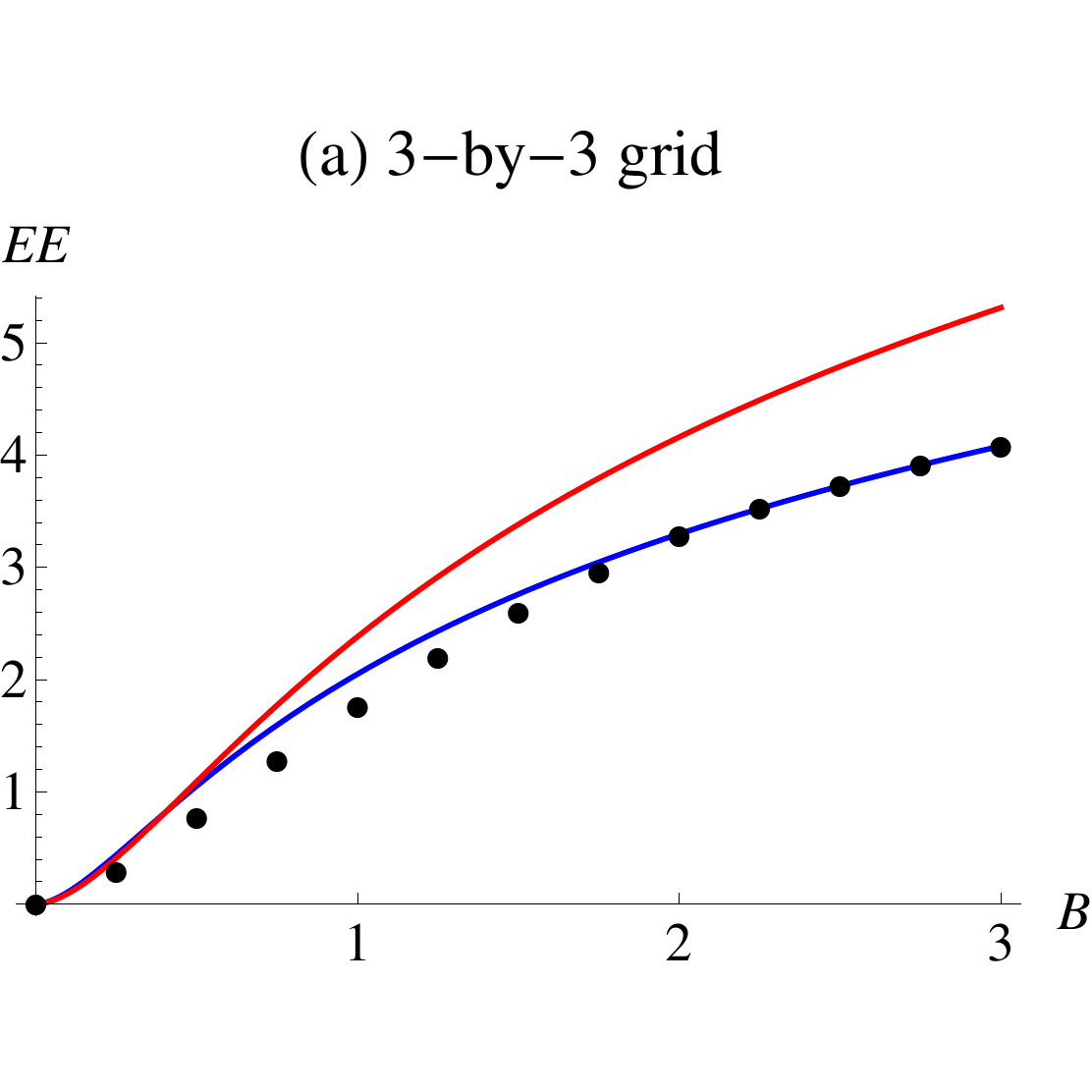}
\includegraphics[width=2in]{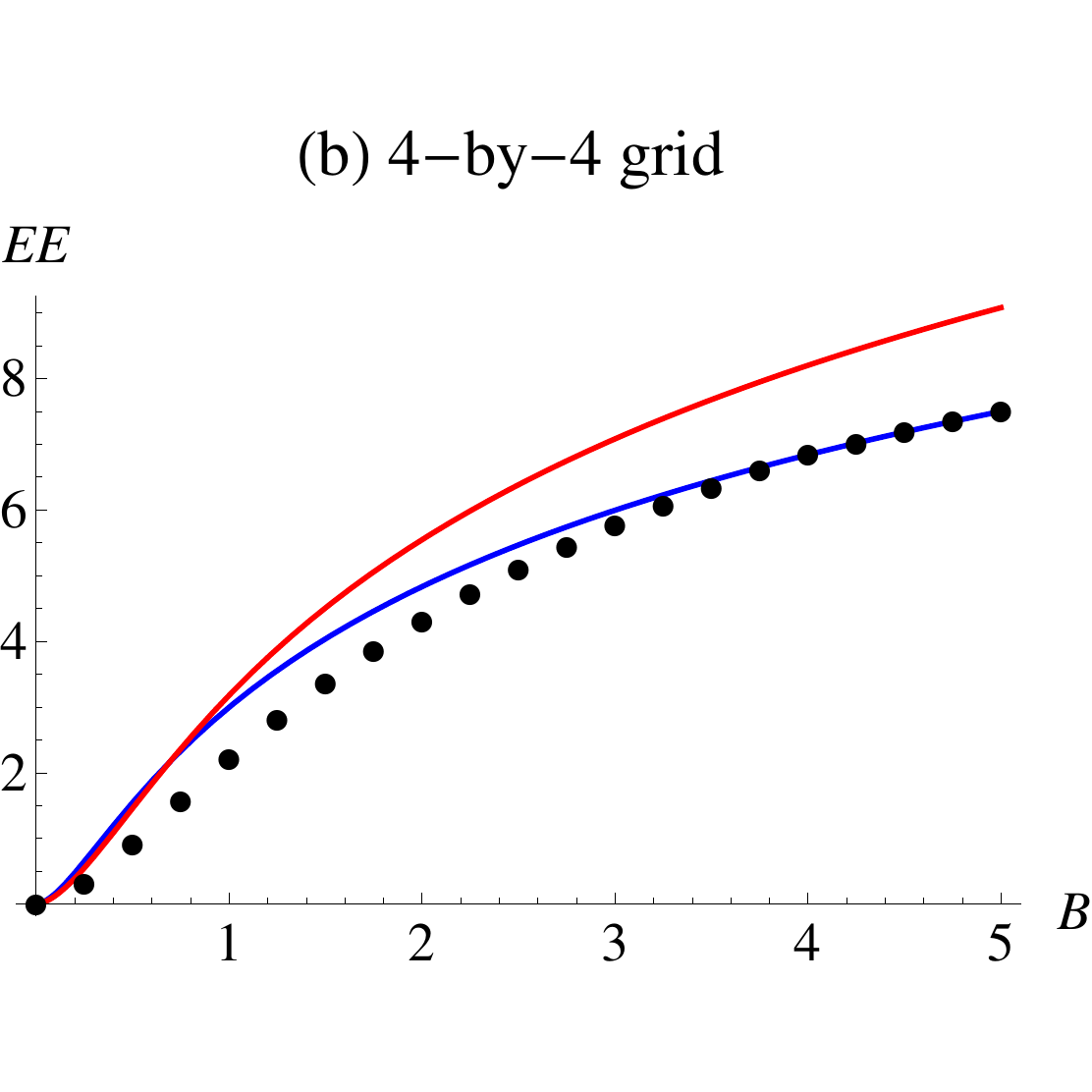}
\end{center}
\caption{\label{fig:NumericalEW} Plots show values of the EW (computed numerically) for a range of the squeezing parameter $B\!=\!\exp\left(-2 \zeta\right)$, compared to the width for the diagonal-type decomposition (blue) and the rectangular-type decomposition (red).  The limit $B\!\longrightarrow\!\infty$, ($\zeta\!\longrightarrow\!-\infty$), corresponds to perfect squeezing for the momentum quadratures for every qumode at the start.  The limit $B\!\longrightarrow\!0$ corresponds to perfect anti-squeezing in the momentum quadratures, and there is no entanglement in this limit.  For both the 3-by-3 lattice, case (a), and the 4-by-4 lattice, case (b), the diagonal-type decomposition is optimal for larger values of $B$ (as would be expected from the qubit case).  However, for this to occur, the required squeezing must be greater for the larger lattice.}
\end{figure}

For any given size of cluster state, we do not know whether either of these two decompositions is optimal, so we have investigated the behaviour for small grids numerically, evaluating the EW for the 3-by-3 and 4-by-4 grids.  To do this, we use an equivalent definition of the EW based on recursion \cite{Oum09}.  Letting $Q$ denote the set of qumode labels, define the function $w(\cdot)$ on all subsets $X$ of $Q$ as:
\begin{eqnarray}
\label{eq:RecursiveEW}
w(X)&=& {\rm min}_{Y\subset X\vert Y\neq \emptyset,Y\neq X} \nonumber \\
&& {\rm max}\{EE(Y,Q\setminus Y),EE(X\setminus Y,Q \setminus (X\setminus Y)),w(Y),w(X \setminus Y)\} \nonumber \\
&& \,\,\,{\rm when} \vert X \vert \geq 2, \nonumber \\
&& EE(X,Q\setminus X)\,\, {\rm when} \vert X \vert\!=\!1.
\end{eqnarray}
Similarly to the previous discussion on decompositions defined by subcubic trees, $w(X)$ can be understood as the minimum width of a binary tree on the subset $X$ of $Q$.  A binary tree has one or three incident edges at every vertex other than the root; removing an edge from the binary tree defines a subset given by the leaves which are the descents of the edge.  The recursive definition here works by combining binary trees which are already optimal on smaller subsets.  The starting point is the values of the EE corresponding to singleton bipartitions of $Q$.  For the case of $X\!=\!Q$, $w(Q)$ now coincides with definition of the EW (an optimal binary tree here be converted into an optimal subcubic tree by removing the root and merging the two incident edges).  (\ref{eq:RecursiveEW}) can be readily implemented using dynamic programming techniques \cite{Wagner95}.  The results are shown in figure~\ref{fig:NumericalEW}.  In both cases, we see that in the limit of high squeezing the EW tends to the diagonal-type decomposition. This behaviour is not surprising, since in the infinite-squeezing limit, the entanglement properties are qubit-like, and the diagonal decomposition is the optimum. We therefore conjecture that, in the limit of high squeezing, the EW of the a general grid will tend to that of the diagonal decomposition, namely (\ref{eq:AnalyDiag}).

\begin{figure}[t]
\begin{center}
\includegraphics[width=2.5in]{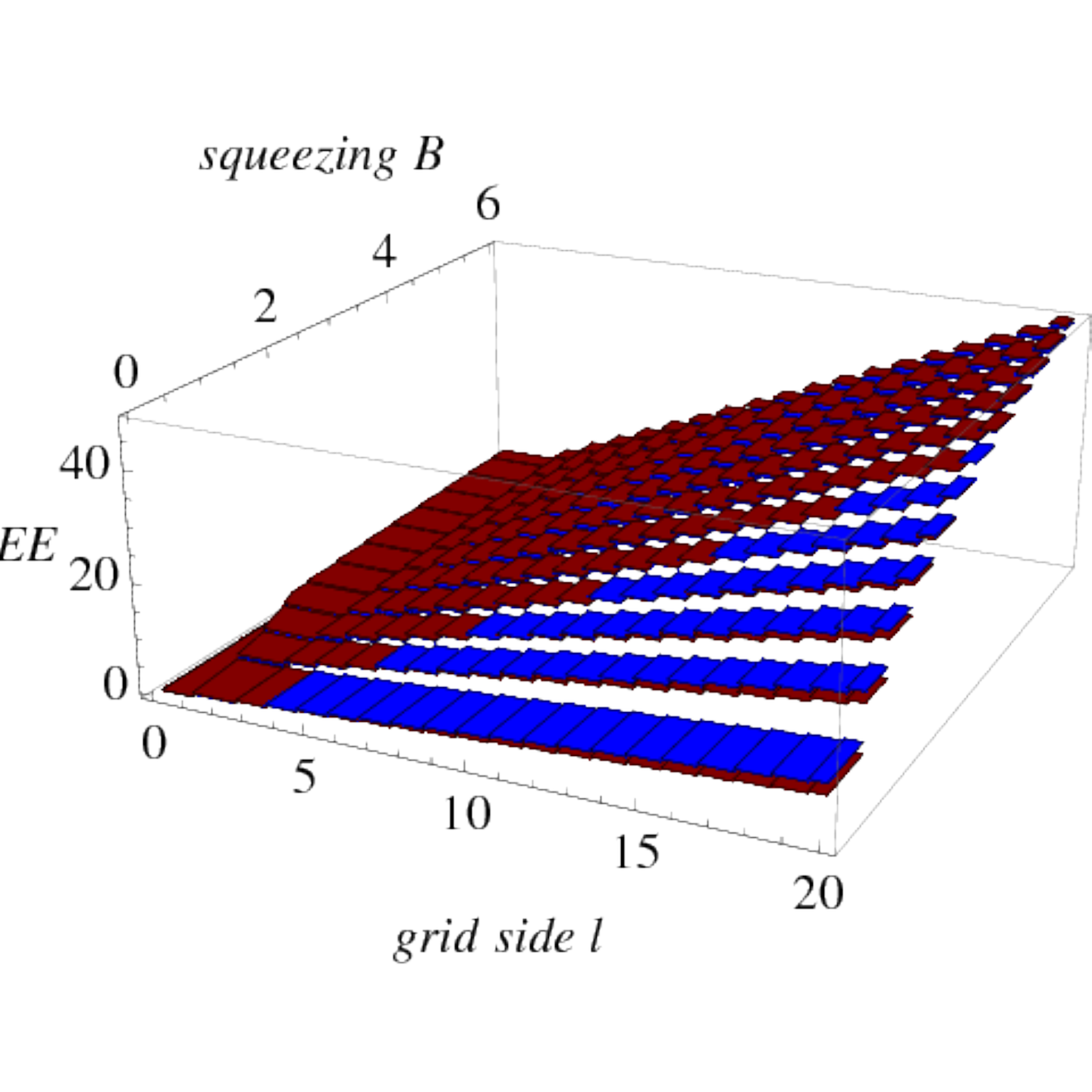}
\end{center}
\caption{\label{fig:DiagVSRect} A comparison of width of the diagonal-type (blue) and rectangle-type (red) decompositions of Gaussian cluster states defined on $l$-by-$l$ grids.  These are determined respectively by the EE of a bipartition defined by a main diagonal, and by a full row (or column) away from the boundary.  Each edge between qumodes, labelled $i$ and $j$, corresponds to the CV controlled-Z operation $\exp(iB\hat{X}_i \hat{X}_j)$.
As $l$ increases with $B$ fixed, the EE for the bipartition defined by the main diagonal becomes greater than that for a bipartition defined by a full row.  In contrast, for the qubit case, the EE is always one less for the diagonal case.
}
\end{figure}

Given that very-high squeezing is currently not experimentally achievable, a more relevant limit is to keep squeezing held constant, and study the behaviour as grid size $l$ gets very large. In this limit, we see a very different behaviour. For fixed $B$, there is always a grid-size $l'$ such that for all grid sizes $l>l'$ the width of the rectangular decomposition is smaller than the diagonal decomposition, and the diagonal decomposition can no longer be optimal. This behaviour can be seen in figure~\ref{fig:DiagVSRect}. The reason for this transition is that both $EE_{\rm diag}(l,B)$ and $EE_{\rm rect}(l,B)$ are linear in $l$, but $\partial EE_{\rm diag}(l,B)/\partial l > \partial EE_{\rm rect}(l,B)/\partial l$. Thus, for a given grid size, something akin to a phase transition occurs in the entanglement properties. We must emphasize though that we cannot know if the  $EE_{\rm rect}(l,B)$ provides a tight upper bound to the EW in this limit. The true EW might be much lower. What this does indicate is that the EW for large, fixed-squeezing grids has a rather different behaviour to either the qubit cluster state or the infinitely-squeezed case.

\section{Bipartite entanglement in the presence of photonic loss}
\label{sec:Loss}

In this section, we consider how our previous results are affected by photonic loss, a principle source of error in optical experiments, especially affecting squeezing operations and storage (necessary for adaptive measurement).  To begin, we consider how a loss channel acts on an arbitrary $N$-mode Gaussian state with covariance matrix $\Gamma_{ij}$ and displacement vector $d_i$. We adopt the standard loss model, for which a mode $r$ couples to an ancilla mode $e$, initially the vacuum, via a beam-splitter-type interaction.  The
beam-splitter transformation acts on the mode operators as
$\hat{a}_{r}\mapsto \sqrt{\eta }\hat{a}_{r}\!+\!\sqrt{1\!-\!\eta }\hat{e}$, with transmissivity parameter $\eta$.  A partial trace is taken for mode $e$ at the end.  This loss channel is Gaussian (that is to say it transforms Gaussian states to Gaussian states), and we write $\Gamma_{ij}^{\rm loss}$ and $d_{i}^{\rm loss}$ for the covariance matrix and displacement vector for the state at the end.  If independent losses are applied to every mode with the same value for $\eta$, the following transformation is readily derived for the phase-space formalism:
\bea
\label{eq:AllModesLossChannel}
\Gamma ^{\rm loss}&=&\eta \Gamma +\left( 1-\eta \right) I, \nonumber \\
{\textbf d}^{\rm loss}&=&\sqrt{\eta}{\textbf d}.
\eea
$I$ here denotes the identity matrix.  The displacement vector can be disregarded, since it corresponds to a local property and plays no role in calculating the entanglement.

The loss channel will be assumed to act on each mode of a Gaussian cluster state after it has been generated from initial squeezing operations on each qumode with parameter $\zeta=-\vert \zeta \vert$, and CV controlled-Z operations between pairs of qumodes.  Squeezing operations do not commute with the loss channel so, unlike the previous entanglement calculations, we can not simply ``undo'' entangling operations within bipartitions to simplify our calculations.  Furthermore, arguments used above for disregarding CV controlled-Z operations within a chosen bipartition, and for combining multiple operations incident at a particular qumode, can no longer be assumed to apply.  To investigate the bipartite entanglement for lossy-Gaussian cluster states, we need an entanglement monotone defined for mixed states.  We choose the logarithmic negativity (LN), defined for a state $\rho$ by $LN\!=\ln \left(\vert\vert\rho^{\rm pt}\vert\vert_1\right)$, where $\vert\vert\cdot\vert\vert_1$ denotes the trace norm and is equal to the sum of the absolute values of the eigenvalues, and ${\rm pt}$ denotes the partial transposition operation on either party \cite{Vidal02}.  As previously, we use the natural logarithm for the CV case.  In operational terms, the LN provides an upper bound on the distillable entanglement.

We can use the LN to define an entanglement width for mixed states.  Applying the central ideas of Van den Nest et al. \cite{VandenNest07}, our new entanglement width LNW must satisfy the following condition: if state $\rho$ with $N$ modes can be converted to state $\rho^\prime$ with $N^\prime\leq N$ modes deterministically, by local operations and classical communication (LOCC), then $LNW(\rho)\geq LNW(\rho^\prime)$.  This condition defines a ``type-II entanglement monotone'', in the terminology of Van den Nest et al.  Differently from those authors, we consider CV states defined on qumodes rather than DV states defined on qubits, and we go beyond the pure state case, defining our type-II monotone on mixed states.  However we retain the two crucial elements of type-II definition: the interconversion by LOCC is assumed to be deterministic and there is no averaging of probabilistic outcomes; states are compared on subsystems of possible different size i.e. the end state can be defined on fewer qumodes.  To verify that in fact our definition of LNW does in fact satisfy the type-II definition, it is sufficient to check two facts.  First, we need to the special case that $LNW(\rho)\geq LNW(\rho^\prime)$ when $N^\prime\!=\!N$.  This follows immediately from the generic min-max construction used to define the entanglement width, since LN is an entanglement monotone in the conventional sense (specifically that it is non-increasing on average under non-deterministic LOCC) \cite{Plenio05}.  Second, we need to check that the LNW is invariant under the addition of an uncorrelated additional party i.e. $LNW(\rho)\otimes LNW(\rho\otimes\ket{0}\!\bra{0}_{N\!+\!1})$.  This is easily verified since, for an arbitrary bipartition, the value of the LN is unchanged by the addition of an extra party.

We thus define the logarithmic-negativity width (LNW) as follows:
\be
LNW(\ket{\rho})={\min_T} \left( {\max_{e\in T}LN}(\beta_e,Q\setminus \beta_e)\right).
\ee
where the width of a decomposition $T$ is defined as the maximum value of the LN for bipartitions of the set of qumode labels $Q$, each one defined by removing an edge $e$ of $T$.  Again the corresponding width, LNW, is defined as the minimum width over all possible decompositions.

We now compute the LNW for a lossy-Gaussian cluster state in a $l$-by-$l$--grid configuration using the phase-space formalism.  As previously, we use $B=\exp(-2\zeta)$ as the relevant squeezing parameter.  Letting $\Gamma^{\rm loss}_{\rm grid}$ denote the covariance matrix for the state, we write:
\bea
\label{eq:CMLossyGrid}
\Gamma^{\rm loss}_{\rm grid}
&=& \eta \left\{
\left[ \bigotimes_{\rm edges\{i,j\}}S_{cZ}^{(i,j)}\right]
\left[ \bigoplus_{q=1}^{l^{2}}
\left(
\begin{array}{cc}
B & 0  \\
0 & 1/B \\
\end{array}
\right)
\right]
\left[ \bigotimes_{\rm edges \{i,j\}}S_{cZ}^{(i,j)}\right] ^{t}\right\} \nonumber \\
&+& (1-\eta )I.
\eea
$S_{cZ}^{(i,j)}$ denotes a CV controlled-Z operation with qumodes with labels $i$ and $j$, and has the value 1 on the main diagonal and at $(i\!+\!l^2,j)$ and $(j\!+\!l^2,i)$.  In the phase-space formalism, the partial transposition operation acts to change the sign of the corresponding momenta.  Given a bipartition of the qumode labels $(\beta_1,\beta_2)$, the operator after partial transposition (over $\beta_1$ say) is of the same form as a Gaussian density matrix with covariance matrix $\Gamma^{\rm loss,PT}_{\rm grid}=P \Gamma^{\rm loss}_{\rm grid} P$, where $P$ is a diagonal matrix having entries $-1$ at positions $\{(q\!+\!l^2,q\!+\!l^2)\vert q\in \beta_1\}$, and value $1$ for the remaining diagonal entries.  Williamson's theorem can be applied to
provide a global symplectic diagonalisation of $\Gamma^{\rm loss,PT}_{\rm grid}$ in terms of symplectic eigenvalues
$\lambda^{\rm PT}_1,\cdots,\lambda^{\rm PT}_{l^2}$.  Modes with symplectic eigenvalues $\lambda^{\rm PT}_i\geq1$ have value 1 for the trace-norm and do not contribute to the LN.  Cases with $0<\lambda^{\rm PT}_i<1$ contradict the uncertainty relation $\Delta \hat{X}\Delta \hat{P}\geq1/2$, and contribute $1/\lambda^{\rm PT}_i$ to the trace norm \cite{Vidal02}.  Overall:
\be
\label{eq:LNPhaseSpace}
LN =\sum^{l^2}_{i=1}{\rm max} \left[-\ln \left( \lambda_i^{\rm PT}\right) ,0\right].
\ee

\begin{figure}[t!]
\begin{center}
\includegraphics[width=1.94in]{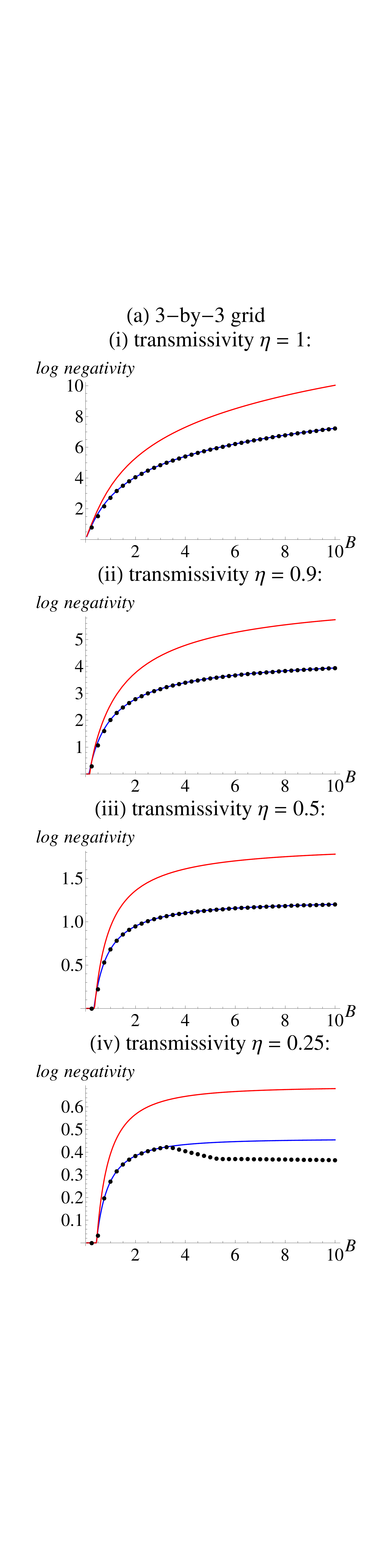}
\includegraphics[width=1.94in]{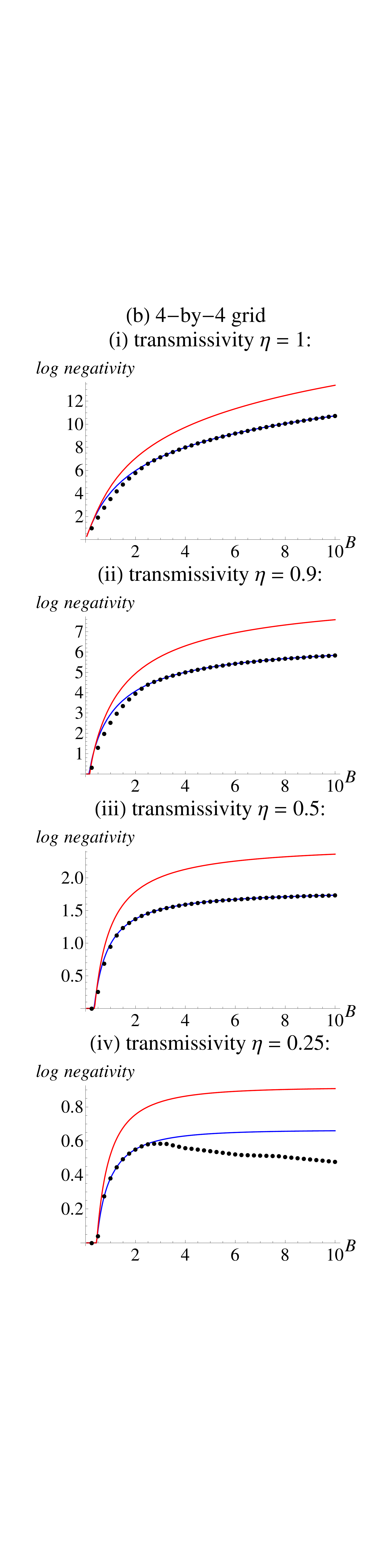}
\end{center}
\caption{\label{fig:LossyLNW}  The LNW is plotted (dots) versus the squeezing parameter $B$ for the (a) 3-by-3 and (b) 4-by-4 grid configurations.  Each subplot corresponds to a different value for transmissivity parameter $\eta$ ($\eta\!=\!1$ is the lossless case).
Blue and red lines are the LN for the diagonal and rectangular decompositions respectively.
}
\end{figure}
In figure \ref{fig:LossyLNW}, Gaussian cluster states in the 3-by-3 and 4-by-4 grid configurations are compared.  When loss is absent or not too large, (i.e. transmissivity $\eta\!=\!0.9$ or 0.5), the diagonal decomposition is optimal, or close-to-optimal across the range of squeezing investigated.  The convergence of the width of the diagonal decomposition to the LNW takes slightly longer for the larger grid but still occurs rapidly.  However, there is a large drop entanglement as loss is added --- for example when $10\%$ loss is added the LNW falls to roughly half the value of the non-dissipative case across the range of squeezing considered.  For the case of extreme loss, for which transmissivity $\eta\!=\!0.25$, the diagonal and rectangular decompositions are seen to be suboptimal for larger values of the squeezing parameter $B$, indicating a change in entanglement structure compared to the idealized case (of infinitely-squeezed CV cluster states or qubit cluster states).  It is seen also that the entanglement grows slowly for larger values of $B$ when loss is added, compared to the non-dissipative case.  This suggests that errors arising from finite squeezing cannot simply be countered by achieving strong squeezing at the start; the infinite-squeezing limit cannot be assumed to behave as in the non-dissipative case.

\begin{figure}[t]
\begin{center}
\includegraphics[width=1.95in]{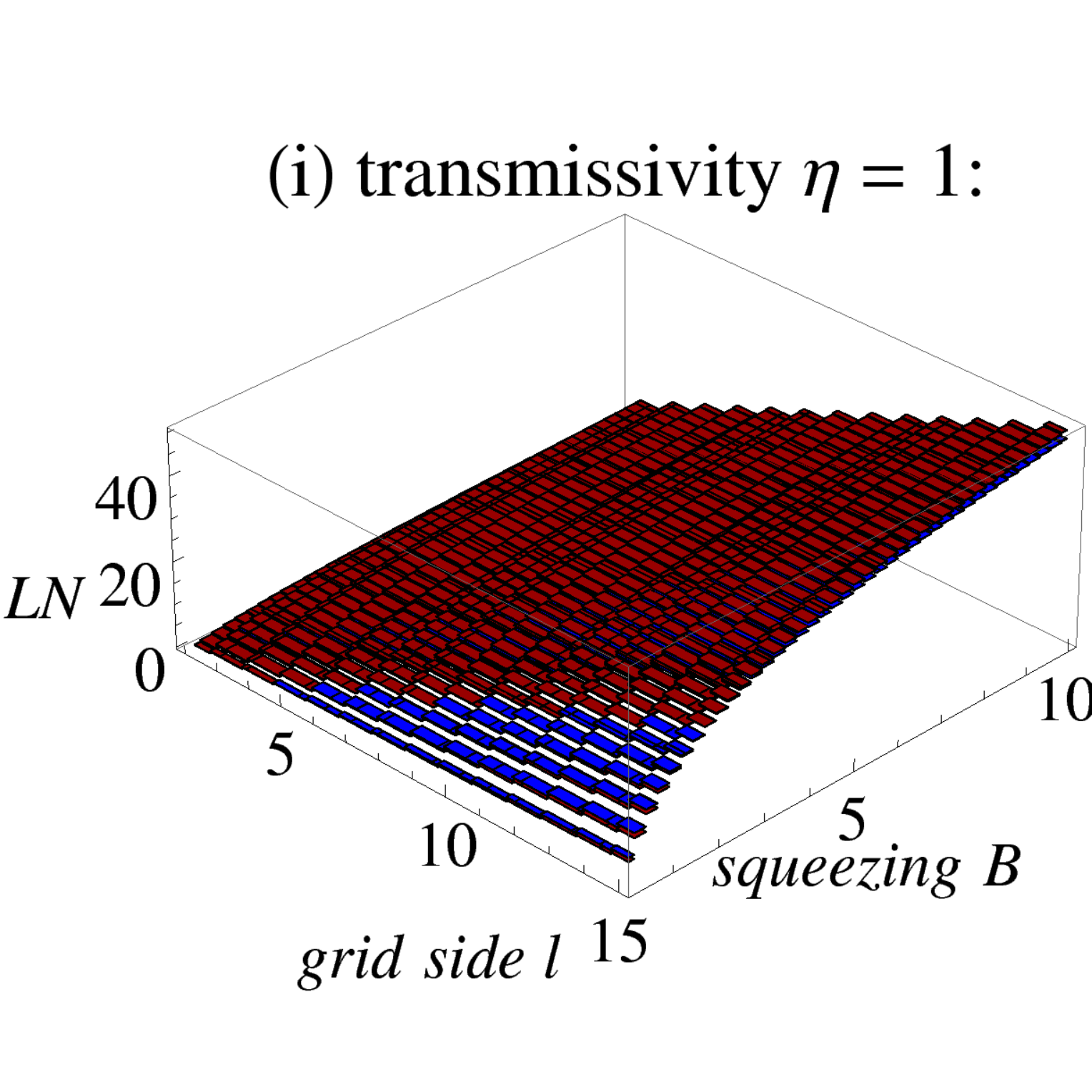}
\includegraphics[width=1.95in]{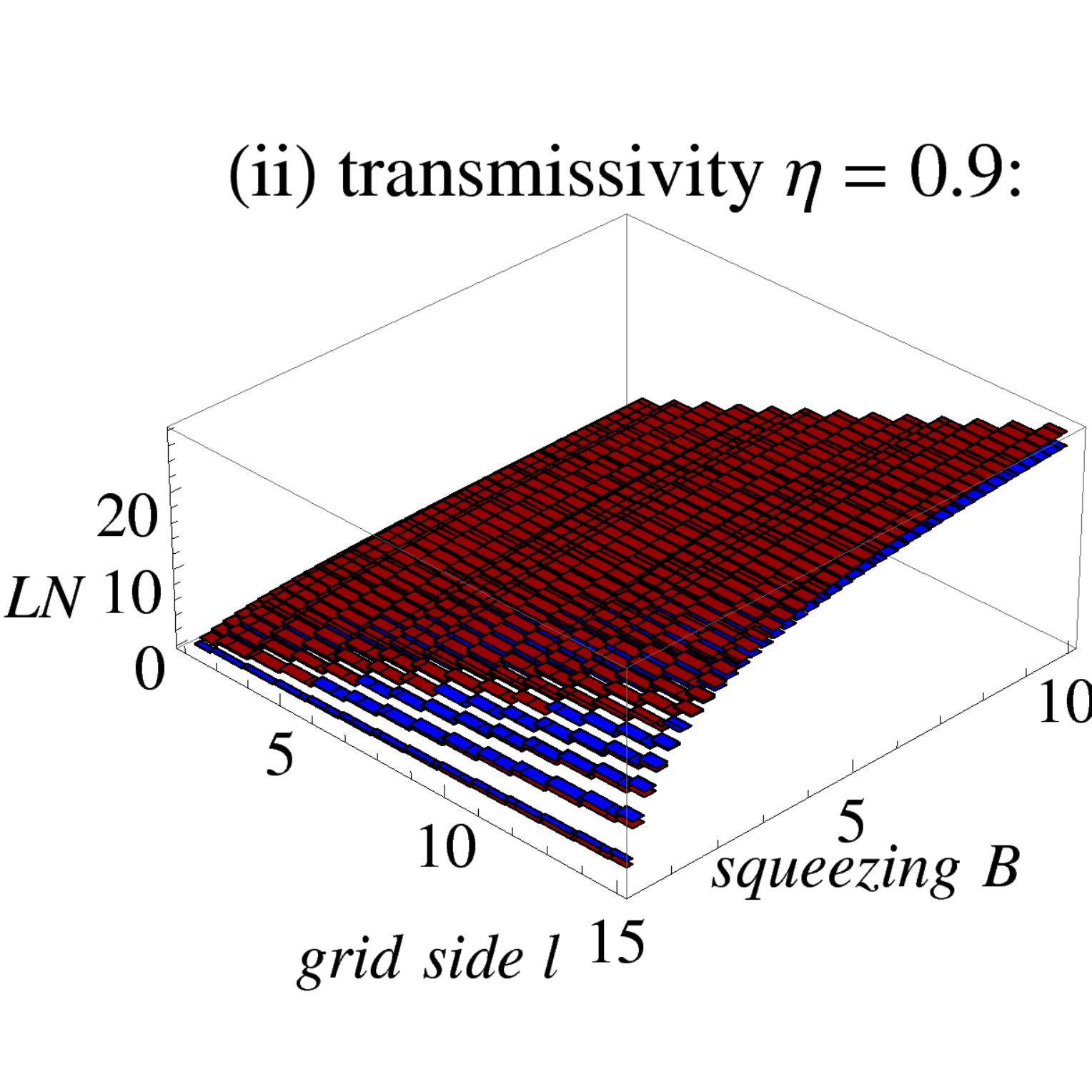}
\includegraphics[width=1.95in]{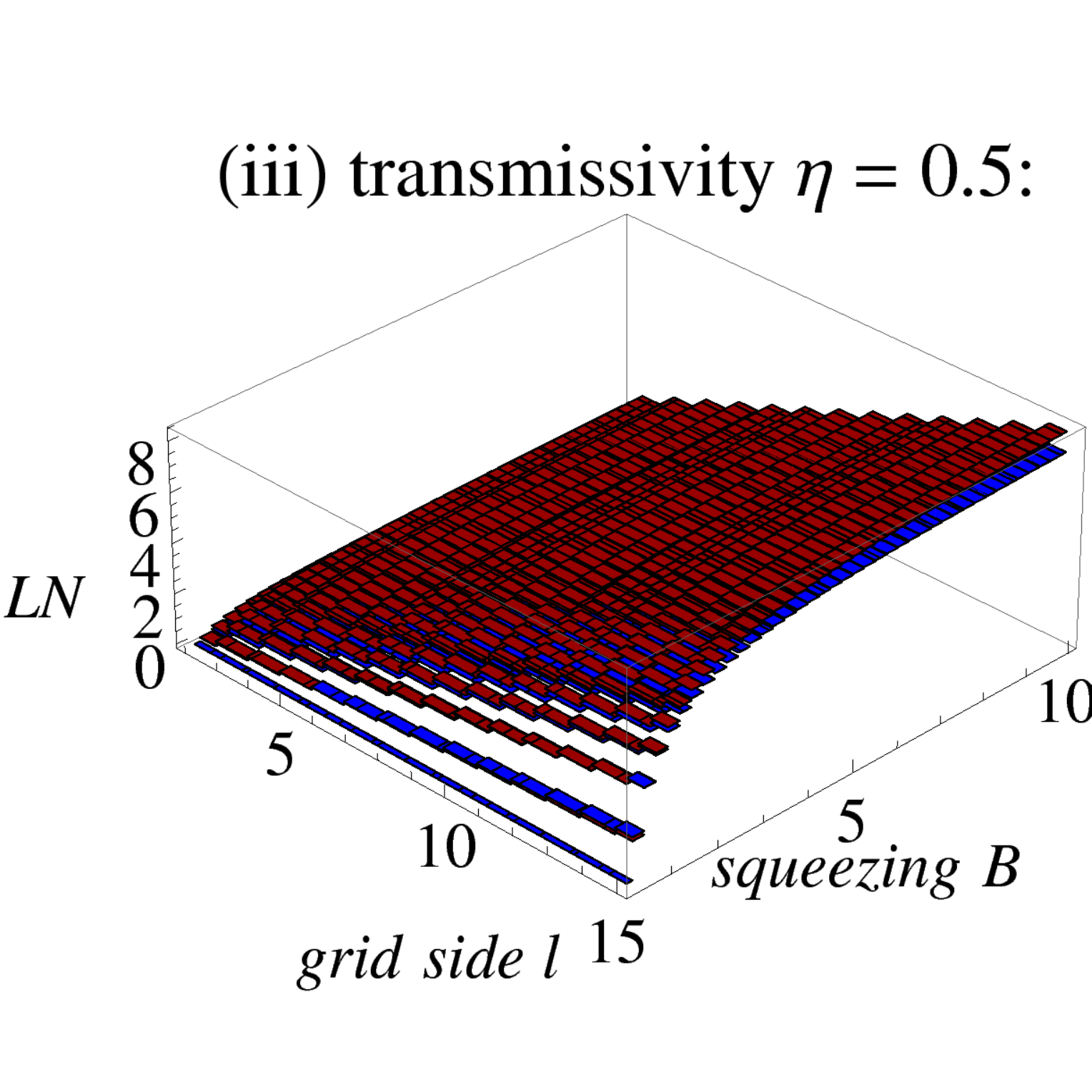}
\end{center}
\caption{\label{fig:LossyDiagVRect} The LN is plotted for the widths of diagonal (blue) and rectangular (red) decompositions for $l$-by-$l$ grid configurations and for a range of the squeezing parameter $B$.  Each subplot corresponds to a different value for transmissivity parameter $\eta$ ($\eta\!=\!1$ is the lossless case).
}
\end{figure}
Finally, the diagonal and rectangular decompositions are compared in figure \ref{fig:LossyDiagVRect} for $l$-by-$l$ grid configurations up to  $l\!=\!15$.  For the non-dissipative case (i), $\eta\!=\!1$, similar behaviour is seen using the LN as a measure of entanglement as was previously observed for the EE (figure \ref{fig:DiagVSRect}).  Specifically, for a fixed value of the squeezed parameter $B$ but increasing grid size, the diagonal decomposition eventually has more entanglement than the rectangular decomposition.  The phenomena becomes less marked as loss is added (cases (ii) transmissivity $\eta\!=\!0.9$ and (iii) $\eta\!=\!0.5$).  In the lossy cases, the key features are the large drop in entanglement (as observed previously for the 3-by-3 and 4-by-4 grids), and the linear growth of the LNW with increasing values $l$ (with $B$ fixed).

\section{Discussion and conclusions}
\label{sec:Conclusions}

In summary, we have investigated the bipartite entanglement properties of Gaussian cluster states under finite squeezing and photonic loss, focusing on quantum wires and grid lattices. In the case of finite squeezing but no loss, we have derived analytic upper bounds for the entanglement (quantified using the EW), which we conjecture to be tight in the limit of high squeezing. We have also shown that a different limiting behaviour occurs in the limit of large grid size, when the initial per-site squeezing $\zeta$ is held fixed, compared to the DV case.  When loss is added to the picture, we have found numerically a large decrease in the available entanglement (quantified using the LNW for mixed states).  Furthermore, we have provided numerical evidence that the LNW for a grid cluster state grows only very slowly as the squeezing parameter $B\!=\!\exp(-2\zeta)$ is increased to large values whenever losses are present.  In all cases however, analytic and numeric evidence point to a linear increase in the entanglement for $l$-by-$l$ grid states with respect to the grid-side $l$ ($B$ held fixed).

It is worth considering the implications of these results for the prospects of scalable measurement-based quantum computation (MBQC) using Gaussian cluster states as a resource. In \cite{VandenNest07}, Van den Nest et al.\ identified a set of criteria which every DV state must satisfy in order to be a ``universal state preparator'' (USP) under local measurements. This requirement, which is stronger than the weakest definition of a universality, (which would only specify a quantum processor providing classical statistical data at the output), is nevertheless satisfied by most known variants of MBQC \cite{Cai05}. The criteria are derived from type-II entanglement monotones, entanglement measures which are non-increasing under deterministic LOCC, where subsystems of possibly-different system size are compared. Every type-II entanglement monotone provides a no-go theorem which can exclude certain families of states as efficient USPs  (for example see theorem 9 in \cite{VandenNest07}).  Variants of the entanglement width, defined in this paper in terms of the entropy of entanglement and the logarithmic negativity, constitute type-II monotones, and we can apply similar arguments about USPs to Gaussian cluster states.\footnote{See \cite{Ohliger10} for an analysis of the Gaussian localisable entanglement in these states.}

The Van den Nest et al.\ criteria are derived by reference to a known USP resource state, typically a grid (or rectangular) cluster state. The essence of the argument can be paraphrased as follows (for a formal description see \cite{VandenNest07}). An efficient USP must be able to generate (efficiently) any state creatable (efficiently) via a quantum circuit. Since this includes the family of grid cluster states itself, this implies necessary criteria relating to the scaling of their entanglement properties.  For example, the EW for any efficient USP resource state must scale faster than logarithmically in the system size.  We would wish to apply a similar criteria to Gaussian cluster states.  A first choice which one must make is whether to compare these potential resource states with a known universal qubit resource (for example the same grid cluster states) or a suitable CV state. Choosing an appropriate CV state is problematic since, so far, the only generally-accepted USP for the CV case is the infinitely-squeezed cluster state. This state (and the high-squeezing limit of finitely-squeezed Gaussian states) cannot be used for comparison, since the entanglement can be unbounded even for the two-mode case.\footnote{A better family of comparison states might be the states of maximal entanglement width which can be generated via a polynomial number of gates (of fixed squeezing). We leave that comparison for future work.} Instead, we shall make our comparison with qubit cluster states, and use directly the criteria derived in \cite{VandenNest07}, namely that the entanglement width must be unbounded (for universality) and must scale faster than logarithmically (for efficient universality).

For the pure-Gaussian cluster states the diagonal and rectangular decompositions both provide an upper bound to the entanglement width. As can be seen from (\ref{eq:AnalyDiag}) and (\ref{eq:AnalyRect}) the entanglement in both these cases scales linearly in grid size, satisfying the criteria for (in principle) efficient conversion to qubit cluster states. These measures only provide upper bounds to the entanglement width, as we have not proven that these bounds are tight.  They do, however, allow us to extract some indication of the capability of these states to support MBQC. In order to create a $l$-by-$l$ qubit cluster state, the Gaussian resource state would have to possess an entanglement width of at least $(l\!-\!1)\ln(2)$. This means that the prefactor in the linear EW scaling rules for the Gaussian cluster states can be thought of as a  ``best-case conversion rate'', quantifying the largest qubit cluster state which could be created via local measurements from a given Gaussian cluster state. For example, the central diagonal bipartition gives us, via (\ref{eq:AnalyDiag}), an upper bound on the conversion rate of $\ln\left(\sqrt{1\!+\!4B^2\!+\!\sqrt{1\!+\!8B^2}}\right)/\ln(2)$. We find that a squeezing factor of $B\approx 0.54$ (corresponding to some anti-squeezing of the momentum quadrature) provides a conversion rate of approximately unity. Note that even a state constructed from initially unsqueezed modes undergoing a full QND interaction would achieve this, and such states may be feasible to construct via a network of beam-splitters acting on single-mode squeezed states of currently achievable squeezing \cite{vanLoock07}. We emphasise, however, that this is a necessary condition only. It does not guarantee that such interconversion is possible, and certainly not that the measurements needed to do so could be achieved via linear optics.

To conclude, while the results of our investigation have led us to less pessimistic conclusions than other recent investigations into Gaussian cluster state entanglement, these results taken together with \cite{Ohliger10} do suggest that utilising Gaussian cluster states for MBQC may require an approach radically different to the traditional qubit approaches \cite{Raussendorf01,Gross08}.  Furthermore, they remind us that no scalable scheme for MBQC with finitely-squeezed cluster states has been proposed so far. We have seen for the entanglement width that the minimizing decomposition, in the limit of large grids, is not the same as the the optimal decomposition in the limit of infinite squeezing.  This indicates that some aspects of the entanglement for finitely-squeezed states are qualitatively different to the infinitely-squeezed limit. The results of section~\ref{sec:Loss} show us that the entanglement width degrades swiftly in the presence of loss, and that coding for tolerance of loss errors will need to be an important component of such schemes.  On the other hand, our results show that, with only a modest initial squeezing, it may be possible to generate states with a greater entanglement-width scaling than for qubit cluster states, and therefore that Gaussian cluster states may potentially be rich resources for entanglement-based protocols. With the impressive recent progress in the experimental generation of Gaussian cluster states \cite{CVClusterExperiments}, developing means to exploit these states for information-processing tasks remains a worthy goal, which deserves continued interest.

\section*{Acknowledgements}

The authors thank Vit Jel{\'i}nek, Mile Gu, Maarten Van den Nest
and Alessio Serafini for helpful discussions.
H.C. acknowledges support for this work by the National
Research Foundation and Ministry of Education, Singapore.
D.E.B. acknowledges financial support from the Leverhulme Trust.

\end{document}